\documentclass[showpacs,10pt,twocolumn,prb]{revtex4-1}

\usepackage{amsmath}
\usepackage{amssymb}
\usepackage{graphicx}
\usepackage{amssymb}
\usepackage{graphics}
\usepackage{epsfig}
\usepackage{CJK}
\usepackage{color}

\begin{document}

\begin{CJK*}{GBK}{}
\title{Polaronic transport and thermoelectricity in Fe$_{1-x}$Co$_x$Sb$_2$S$_4$ ($x$ = 0, 0.1, and 0.2)}
\author{Yu Liu,$^{1}$ Chang-Jong Kang,$^{2}$ Eli Stavitski,$^{3}$, Qianheng Du,$^{1,4}$  Klaus Attenkofer,$^{3}$ G. Kotliar$^{1,2}$ and C. Petrovic$^{1,4}$}
\affiliation{$^{1}$Condensed Matter Physics and Materials Science Department, Brookhaven National Laboratory, Upton, New York 11973, USA\\
$^{2}$Department of Physics and Astronomy, Rutgers University, Piscataway, New Jersey 08856, USA\\
$^{3}$National Synchrotron Light Source II, Brookhaven National Laboratory, Upton, New York 11973, USA\\
$^{4}$Department of Materials Science and Chemical Engineering, Stony Brook University, Stony Brook, New York 11790, USA
}
\date{\today}

\begin{abstract}
We report a study of Co-doped berthierite Fe$_{1-x}$Co$_x$Sb$_2$S$_4$ ($x$ = 0, 0.1, and 0.2). The alloy series of Fe$_{1-x}$Co$_x$Sb$_2$S$_4$ crystallize in an orthorhombic structure with the Pnma space group, similar to FeSb$_2$, and show semiconducting behavior. The large discrepancy between activation energy for conductivity, $E_\rho$ (146 $\sim$ 270 meV), and thermopower, $E_S$ (47 $\sim$ 108 meV), indicates the polaronic transport mechanism. Bulk magnetization and heat capacity measurements of pure FeSb$_2$S$_4$ ($x$ = 0) exhibit a broad antiferromagnetic (AFM) transition ($T_N$ = 46 K) followed by an additional weak transition ($T^*$ = 50 K). Transition temperatures ($T_N$ and $T^*$) slightly decrease with increasing Co content $x$. This is also reflected in the thermal conductivity measurement, indicating strong spin-lattice coupling. Fe$_{1-x}$Co$_x$Sb$_2$S$_4$ shows relatively high value of thermopower (up to $\sim$ 624 $\mu$V K$^{-1}$ at 300 K) and thermal conductivity much lower when compared to FeSb$_{2}$, a feature desired for potential applications based on FeSb$_{2}$ materials.
\end{abstract}

\maketitle
\end{CJK*}

\section{INTRODUCTION}

Correlated electron materials may enable transformative changes in thermoelectric energy creation and conversion. The Kondo-insulator-like semiconductor FeSb$_2$ features not only strong electronic correlations but also the highest thermoelectric power factor in nature and thermopower up to 45 mV K$^{-1}$.\cite{Petrovic1, Petrovic2, HuR2, Bentien, Jie, Takahashi} For predictive theory modeling of correlated electron thermoelectricity, a similar chemically tunable material is of high interest.

The ternary MPn$_2$Q$_4$ (M = Mn, Fe; Pn = Sb, Bi; Q = S, Se) compounds are typically magnetic semiconductors that exhibit high thermopower and rather tunable electronic, magnetic, and thermoelectric properties.\cite{Tian, Leone, Matar, Pfitzner, Djieutedjeu1, Djieutedjeu2, Li, Lee, Ranmohotti1, Lukaszewicz, Wintenberger, Bonville, Periotto, Djieutedjeu3, Ranmohotti2} For instance, antiferromagnetic (AFM) ordering can be observed in FeSb$_2$S$_4$,\cite{Lukaszewicz, Wintenberger, Bonville, Periotto} MnSb$_2$S$_4$,\cite{Tian, Leone, Matar, Pfitzner} MnSb$_2$Se$_4$,\cite{Djieutedjeu1, Djieutedjeu2, Li} and MnBi$_2$Se$_4$,\cite{Lee, Ranmohotti1} whereas FeSb$_2$Se$_4$ and FeBi$_2$Se$_4$ exhibit ferromagnetic (FM) behavior.\cite{Djieutedjeu3, Ranmohotti2} Moreover, p-type semiconducting behavior is observed in MnSb$_2$Se$_4$ and FeSb$_2$Se$_4$ with a semiconductor-to-insulator transition for FeSb$_2$Se$_4$ below 130 K,\cite{Djieutedjeu1, Djieutedjeu2, Li, Djieutedjeu3} but MnBi$_2$Se$_4$ and FeBi$_2$Se$_4$ are n-type semiconductors.\cite{Lee, Ranmohotti1, Ranmohotti2}

Among these compounds, FeSb$_2$S$_4$ shows a helicoidal-type AFM order below $T_N$ = 50 K with Fe$^{2+}$ moments parallel to the $ab$ plane and a non-commensurate propagation vector along the $c$ axis.\cite{Bonville} The unit cell contains four FeSb$_2$S$_4$, in which Fe atoms are surrounded by six S atoms in a distorted octahedral arrangement. The FeS$_6$ octahedra share edges to form chains parallel to the $b$ axis [Fig. 1(a)], which is similar to FeSb$_2$.\cite{Hulliger1,Hulliger2,Goodenough1} The chains are connected together via S-Sb-S bonds with some rather short Sb-S distances (2.43 {\AA} and 2.48 {\AA}) suggesting strong covalence of these bonds, whereas the large Fe-S distances (2.45 $\sim$ 2.62 {\AA}) indicate that the Fe-S bond is rather ionic and that Fe is in the 3$d_6$ high spin Fe$^{2+}$ state.\cite{Lukaszewicz, Bonville} Furthermore, FeSb$_2$S$_4$ features a lone Sb$^{3+}$ pair which could increase anharmonicity of bonds and enhance phonon-phonon scattering.\cite{Periotto} In contrast to literature devoted to MnPn$_{2}$Q$_{4}$ or FeSb$_{2}$Se$_{4}$, there are no studies of FeSb$_2$S$_4$ thermoelectric and/or physical properties tuning yet. The evolution of crystal structure tuned by 50 \% Bi or Nd doped at Sb site was studied without physical properties measurements.\cite{Bindi, Gasymov}

Here, we investigate a series of Co-doped berthierite Fe$_{1-x}$Co$_{x}$Sb$_2$S$_4$ ($x$ = 0, 0.1, and 0.2). In contrast to Fe$_{1-x}$Co$_{x}$Sb$_{2}$ and Fe$_{1-x}$Cr$_{x}$Sb$_{2}$ where electronic transport is dominated by thermal activation and variable range hopping,\cite{HuR3,HuR4} our results indicate polaronic transport and strong spin-lattice coupling. Higher ionicity of chemical bonds in Fe-S octahedra when compared to Fe-Sb octahedra inhibits electrical conductivity. In contrast to FeSb$_{2}$, however, in these materials we report lower thermal conductivity when compared to FeSb$_{2}$ due to induced phonon-scattering lattice distortions and disorder introduced by Co atoms.

\section{EXPERIMENTAL DETAILS}

Fe$_{1-x}$Co$_x$Sb$_2$S$_4$ polycrystals were synthesized via solid state reaction starting from an intimate mixture of high purity elements Fe powder (99.99 $\%$, Alfa Aesar), Co powder (99.99 $\%$, Alfa Aesar), Sb pieces (99.999 $\%$, Alfa Aesar), and S powder (99.9 $\%$, Alfa Aesar) with a molar ratio of $1-x$ : $x$ : 2 : 4. The starting materials were mixed and ground in an agate mortar, then pressed into pellets and sealed in an evacuated quartz tube backfilled with pure argon gas. The tube was heated to 300 $^\circ$C over 10 h, held at 300 $^\circ$C for 10 h, and then slowly heated to 500 $^\circ$C and reacted for 5 days followed by furnace cooling. This procedure was repeated several times to ensure homogeneity. Powder x-ray diffraction (XRD) data were taken with Cu K$_{\alpha}$ ($\lambda=0.15418$ nm) radiation of Rigaku Miniflex powder diffractometer. The structural parameters were obtained by Rietveld refinement using RIETICA software. X-ray absorption spectroscopy was measured at 8-ID beamline of the National Synchrotron Light Source II (NSLS II) at Brookhaven National Laboratory (BNL) in the transmission mode. The extracted extended x-ray absorption fine structure (EXAFS) signal, $\chi(k)$, was weighed by $k^2$ to emphasize the high-energy oscillation and then Fourier-transformed in a $k$ range from 2.5 to 8.5 {\AA}$^{-1}$ to analyze the data in $R$ space. Thermal, transport, and magnetic measurements were carried out in the Quantum Design PPMS-9 and MPMS-5 systems. The electronic structure of the non-magnetic FeSb$_{2}$S$_{4}$ is calculated within the full-potential linearized augmented plane wave (LAPW) method implemented in WIEN2k package.\cite{Weinert,Blaha} The general gradient approximation (GGA) was used for exchange-correlation potential.\cite{Perdew} The Brillouin zone is sampled with a Gamma-centered 12$\times$38$\times$10 k-space mesh.

\section{RESULTS AND DISCUSSIONS}

\begin{figure}
\centerline{\includegraphics[scale=1]{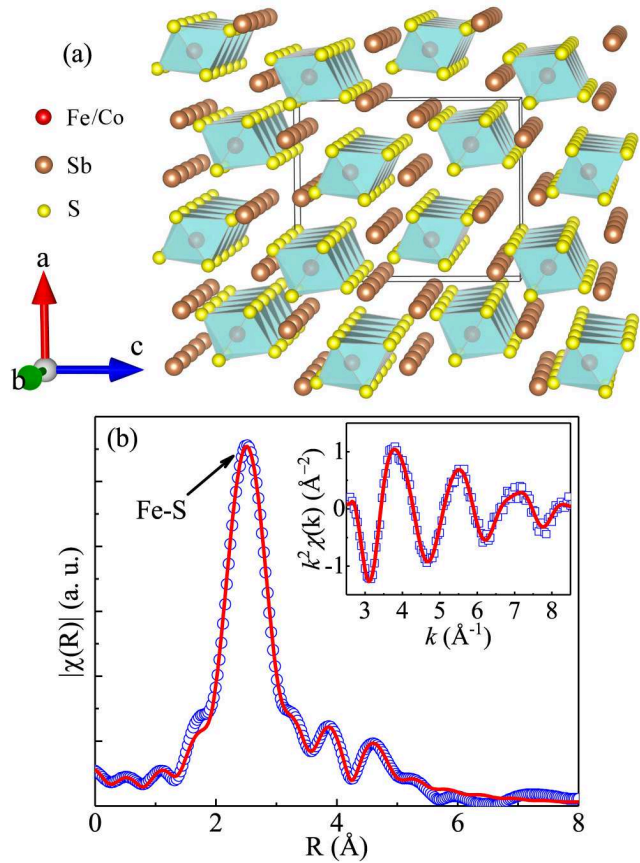}}
\caption{(Color online). (a) Crystal structure. (b) Fourier transform magnitudes of the extended x-ray absorption fine structure (EXAFS) data of FeSb$_2$S$_4$ measured at room temperature. The experimental data are shown as blue symbols alongside the model fit plotted as red line. Inset in (b) shows the corresponding EXAFS oscillation with the model fit.}
\label{XRD}
\end{figure}

The crystal structure of FeSb$_2$S$_4$ was first determined by Buerger \emph{et al.}.\cite{Buerger} The structure contains three distinct cation positions: Fe$^{2+}$ has an octahedral coordination and the polyhedra share edges with two conjugate Sb$^{3+}$ coordination polyhedra.\cite{Periotto} The Fe$^{2+}$ coordination octahedra also share opposite edges among themselves and form chains along [010], similar to FeSb$_2$, but with Sb$^{3+}$ cations inserted between the chains [Fig. 1(a)].\cite{Lukaszewicz} The local structure was determined by EXAFS spectra [Fig. 1(b)] of FeSb$_2$S$_4$ measured at room temperature. In the single-scattering approximation, the EXAFS could be described by the following equation,\cite{Prins}
\begin{align*}
\chi(k) = \sum_i\frac{N_iS_0^2}{kR_i^2}f_i(k,R_i)e^{-\frac{2R_i}{\lambda}}e^{-2k^2\sigma_i^2}sin[2kR_i+\delta_i(k)],
\end{align*}
where $N_i$ is the number of neighbouring atoms at a distance $R_i$ from the photoabsorbing atom. $S_0^2$ is the passive electrons reduction factor, $f_i(k, R_i)$ is the backscattering amplitude, $\lambda$ is the photoelectron mean free path, $\delta_i$ is the phase shift, and $\sigma_i^2$ is the correlated Debye-Waller factor measuring the mean square relative displacement of the photoabsorber-backscatter pairs. In FeSb$_2$S$_4$, the first nearest neighbors of Fe atoms are six S atoms located at 2.45 {\AA} $\sim$ 2.62 {\AA}, and the second nearest neighbors are Fe atoms and Sb atoms at about 3.76 {\AA}.\cite{Lukaszewicz} As shown in Fig. 1(b), the corrected main peak around $R \sim 2.5$ {\AA} in the Fourier transform magnitudes of Fe K-edge EXAFS corresponds to three different Fe-S bond distances with 2.449(2) {\AA}, 2.501(2) {\AA}, and 2.614(2) {\AA} extracted from the model fits with $N$ fixed to 2 and $\sigma^2 = 0.014$ {\AA}$^2$. The peaks between 3.6 {\AA} and 5.0 {\AA} are due to longer Fe-Fe ($\sim$ 3.765 {\AA}) and Fe-Sb ($\sim$ 3.762 {\AA}, 4.000 {\AA}, 4.316 {\AA}, and 4.521 {\AA}) bond distances, and the multiple scattering involving different near neighbours of the Fe atoms. The salient features of the local crystallographic environment of Fe atoms are in good agreement with the previous studies of the average crystal structure.\cite{Buerger}

Figure 2(a) shows the structural refinement of powder XRD for Fe$_{1-x}$Co$_x$Sb$_2$S$_4$ ($x$ = 0, 0.1, and 0.2), indicating that all reflections can be well indexed in the Pnma space group. A tiny peak ($\sim 35^\circ$) of CoSbS emerges as $x$ = 0.2 (less than 5 \%), in line with the absence of a stable phase of CoSb$_2$S$_4$. For pure FeSb$_2$S$_4$, the determined lattice parameters $a$ = 11.385(2) {\AA}, $b$ = 3.765(2) {\AA}, and $c$ = 14.147(2) {\AA}, which are reasonably smaller than those of orthorhombic MnSb$_2$S$_4$ ($a$ = 11.459 {\AA}, $b$ = 3.823 {\AA}, and $c$ = 14.351 {\AA}).\cite{Matar} M$\mathrm{\ddot{o}}$ssbauer spectra and theoretical calculations for FeSb$_2$S$_4$ and MnSb$_2$S$_4$ suggest that Fe$^{2+}$ and Mn$^{2+}$ are in the high spin state.\cite{Matar, Bonville} This is in agreement with  $r_{Fe^{2+}}$ (0.78 {\AA}) $<$ $r_{Mn^{2+}}$ (0.83 {\AA}) for the high spin state with sixfold coordination. Figure 2(b) shows the evolution of lattice parameters with Co doping content $x$, in which the relative change of $a$ slightly increases with $x$ ($\delta_a \approx 0.11\%$), whereas $b$ and $c$ monotonously decrease with $x$ ($\delta_b \approx -0.16\%$ and $\delta_c \approx -0.11\%$).

\begin{figure}
\centerline{\includegraphics[scale=1]{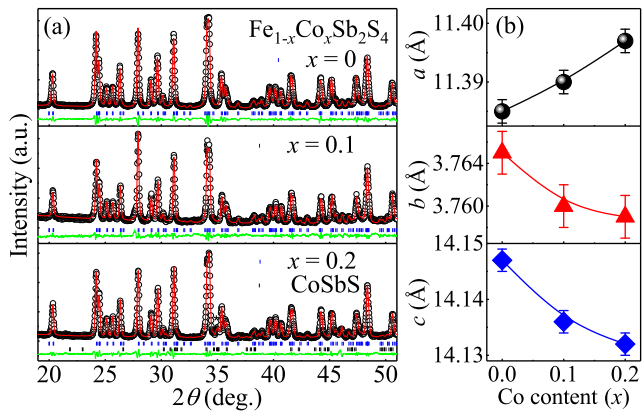}}
\caption{(Color online). (a) Powder x-ray diffraction (XRD) patterns for Fe$_{1-x}$Co$_x$Sb$_2$S$_4$ ($x$ = 0, 0.1, and 0.2). Impurity peak of CoSbS is labeled by an asterisk. (b) The evolution of lattice parameters $a$, $b$, and $c$.}
\label{EXAFS}
\end{figure}

Temperature-dependent electrical resistivity $\rho(T)$ for Fe$_{1-x}$Co$_x$Sb$_2$S$_4$ ($x$ = 0, 0.1, and 0.2) is depicted in Fig. 3(a), showing an obvious semiconducting behavior. The value of room temperature resistivity ($\rho_{300K}$) is about 5.6 $\Omega$ cm for FeSb$_2$S$_4$, which is smaller than the value of 16 $\Omega$ cm for FeSb$_2$Se$_4$,\cite{Djieutedjeu3} and it gradually increases to 35 $\Omega$ cm and 62 $\Omega$ cm for Fe$_{1-x}$Co$_x$Sb$_2$S$_4$ with $x$ = 0.1 and 0.2, respectively. Three typical models are considered to describe the semiconducting behavior: (i) thermally activated model, $\rho(T) = \rho_0 exp(\frac{E_\rho}{k_BT})$, where $E_\rho$ is activation energy; (ii) adiabatic small polaron hopping model, $\rho(T) = AT exp(\frac{E_\rho}{k_BT})$;\cite{Austin} and (iii) Mott's variable-range hopping (VRH) model, $\rho(T) = \rho_0 exp(\frac{T_0}{T})^{1/4}$. To well understand the transport mechanism in this system, it is necessary to fit the resistivity curves based on these three formulas. Figure 3(b) shows the fitting result of the adiabatic small polaron hopping model. The extracted activation energy $E_\rho$ [inset in Fig. 3(b)] is about 146(1) meV for $x$ = 0, and gradually increases to 270(1) meV for $x$ = 0.2. For FeSb$_2$S$_4$, the estimated band gap of about 0.292(2) eV is relatively smaller than the values of FeSb$_2$Se$_4$ (0.33 eV),\cite{Djieutedjeu3} MnSb$_2$S$_4$ (0.77 eV),\cite{Pfitzner} and MnSb$_2$Se$_4$ (0.31 eV).\cite{Djieutedjeu1} In fact, the $\rho(T)$ curves can also be well fitted using the thermally activated model but not the VRH model.

To distinguish the thermally activated model and polaron hopping model, we further measured temperature-dependent thermopower $S(T)$. The $S(T)$ shows positive values in the whole temperature range [Fig. 3(c)], indicating dominant hole-type carriers. In the inset in Fig. 3(c), the room temperature $S_{300K}$ value of FeSb$_2$S$_4$ is about 464 $\mu$V K$^{-1}$. The Co doping at Fe site increases thermopower $S$ and reaches $S_{300K}$ = 624 $\mu$V K$^{-1}$ for Fe$_{1-x}$Co$_x$Sb$_2$S$_4$ with $x$ = 0.1. It gradually increases with the decreasing temperature to a value of 848 $\mu$V K$^{-1}$ at 200 K. As shown in Fig. 3(d), the $S(1000/T)$ curves of all samples show similar shape and can be fitted with the equation $S(T) = \frac{k_B}{e}(\alpha+\frac{E_S}{k_BT})$,\cite{Austin} where $E_S$ is activation energy and $\alpha$ is a constant. The obtained activation energy for thermopower, $E_S$ (47 $\sim$ 108 meV) [inset in Fig. 3(d)] are much smaller than those for conductivity, $E_\rho$ (146 $\sim$ 270 meV) [inset in Fig. 3(b)]. The large discrepancy between $E_S$ and $E_\rho$ typically reflects the polaron transport mechanism of carriers. According to the polaron model, the $E_S$ is the energy required to activate the hopping of carriers, while $E_\rho$ is the sum of the energy needed for the creation of carriers and activating the hopping of carriers.\cite{Austin} Therefore, within the polaron hopping model the activation energy $E_S$ is smaller than $E_\rho$.

\begin{figure}
\centerline{\includegraphics[scale=1]{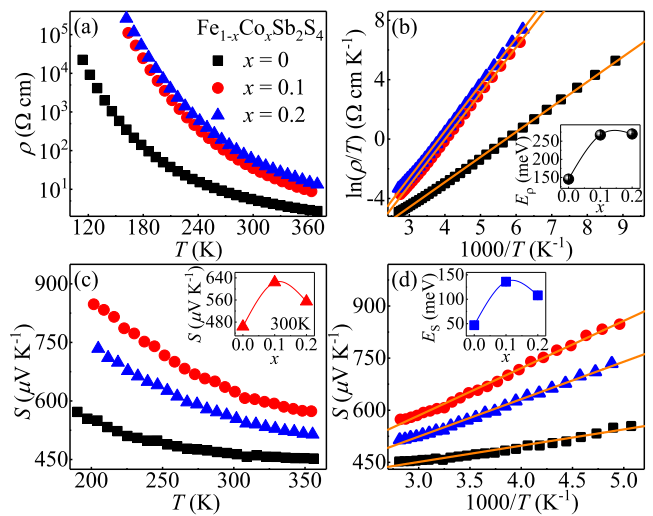}}
\caption{(Color online). (a) Temperature-dependent electrical resistivity $\rho(T)$ of Fe$_{1-x}$Co$_x$Sb$_2$S$_4$ ($x$ = 0, 0.1, and 0.2). (b) ln($\rho/T$) vs 1000/$T$ curves fitted by the adiabatic small polaron hopping model, $\rho(T) = ATexp(E_\rho/k_BT)$, where $E_\rho$ is activation energy and $k_B$ is Boltzmann constant. Inset: The evolution of $E_\rho$. (c) Temperature-dependent thermopower $S(T)$ of Fe$_{1-x}$Co$_x$Sb$_2$S$_4$ ($x$ = 0, 0.1, and 0.2). Inset: the values of thermopower at room temperature. (d) $S(T)$ vs 1000/$T$ curves fitted using $S(T) = (k_B/e)(\alpha+E_S/k_BT)$, where $E_S$ is activation energy. Inset: The evolution of $E_S$.}
\label{RST}
\end{figure}

\begin{figure}
\centerline{\includegraphics[scale=1]{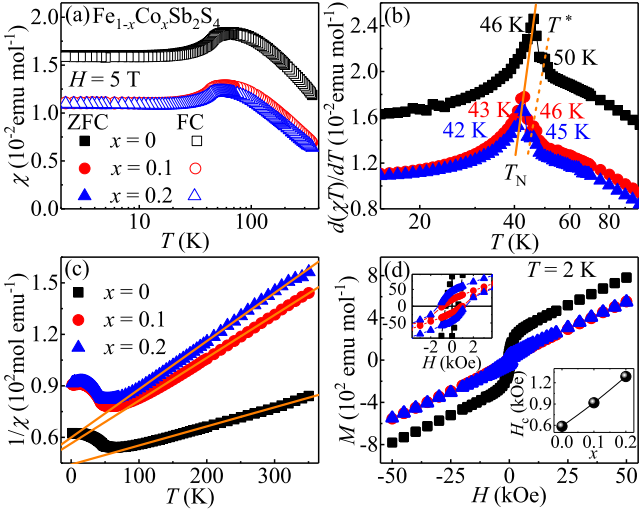}}
\caption{(Color online) (a) Temperature-dependent magnetic susceptibility obtained at $H$ = 5 T with zero-field cooling (ZFC) and field cooling (FC) modes. (b) $1/\chi$ vs $T$ fitted by the modified Curie-Weiss law $\chi = \chi_0 + \frac{C}{T-\theta}$, where $\chi_0$ is the temperature-independent susceptibility, $C$ is the Curie-Weiss constant, and $\theta$ is the Weiss temperature. (c) $d(\chi T)/dT$ vs $T$ curves. The solid and dashed lines are guides to the eye. (d) The hysteresis loops taken at $T$ = 2 K of Fe$_{1-x}$Co$_x$Sb$_2$S$_4$ ($x$ = 0, 0.1, and 0.2). Inset: The magnification in the low field region and the evolution of coercive field $H_c$.}
\label{MTH}
\end{figure}

Temperature dependence of dc magnetic susceptibility $\chi(T) = M/H$ taken in $H$ = 5 T for Fe$_{1-x}$Co$_x$Sb$_2$S$_4$ with zero-field cooling (ZFC) and field cooling (FC) modes are shown in Fig. 4(a). A broad susceptibility maximum around $T$ $\approx$ 50 K was observed in FeSb$_2$S$_4$, in agreement with the previous report.\cite{Wintenberger} It implies that there is a low-dimensional AFM spin correlation among Fe$^{2+}$ ions. As shown in Fig. 4(b), the AFM transition temperatures ($T_N$) of Fe$_{1-x}$Co$_x$Sb$_2$S$_4$ ($x$ = 0, 0.1, and 0.2) are defined by the maxima of $d(\chi T)/dT$ vs $T$ curves. With Co doping, the AFM transition is robust and $T_N$ shows weak shift to lower temperatures. Additionally, no divergence of the ZFC and FC curves was observed in Fe$_{1-x}$Co$_x$Sb$_2$S$_4$, which is different from the other members of MPn$_2$Q$_4$ (M = Mn, Fe; Pn = Sb, Bi; and Q = S, Se) system.\cite{Djieutedjeu1, Djieutedjeu3, Ranmohotti1, Ranmohotti2, Leone} Taken into account the large intralayer distance of $\sim$ 7 {\AA} and the interlayer separation of $\sim$ 15 {\AA} between MQ$_6$ magnetic chains within the crystal structure of MPn$_2$Q$_4$ system, the magnetic properties are mostly controlled by the nature and magnitude of indirect exchange interactions between adjacent magnetic atoms through the bridging Q atoms within the individual MQ$_6$ magnetic chain (intrachain). Within the single chain of FeS$_6$ octahedra, according to the Goodenough-Kanamori rules,\cite{Goodenough2} superexchange interactions at 90$^\circ$ are antiferromagnetic. Furthermore, the $T_N$ of FeSb$_2$S$_4$ ($\sim$ 46 K) is higher than that of MnSb$_2$S$_4$ ($\sim$ 25 K), implying stronger interaction due to smaller Fe-S-Fe distance.\cite{Leone} It is also reflected by the evolution of $T_N$ in Mn-based MnPn$_2$Q$_4$, in which the $T_N$ of MnSb$_2$S$_4$ ($\sim$ 25 K) with smaller Mn-S-Mn distance is higher than those of MnSb$_2$Se$_4$ ($\sim$ 20 K) and MnBi$_2$Se$_4$ ($\sim$ 15 K) with larger Mn-Se-Mn distances.\cite{Djieutedjeu1, Ranmohotti1, Leone} The susceptibility data above 100 K could be well fitted to the modified Curie-Weiss law, $\chi = \chi_0 + \frac{C}{T-\theta}$, where $\chi_0$ is the temperature-independent susceptibility, $C$ is the Curie-Weiss constant, and $\theta$ is the Weiss temperature. As shown in Fig. 4(c), a linear fit of the $1/\chi$ curve of FeSb$_2$S$_4$ yields the Weiss temperature $\theta$ = -397(1) K, confirming predominantly AFM interaction between Fe$^{2+}$ moments. With Co doping, the value of $\theta$ changes to -226(1) K and -213(1) K for Fe$_{1-x}$Co$_x$Sb$_2$S$_4$ with $x$ = 0.1 and 0.2, respectively, indicating weaken AFM interactions. The decrease in the absolute value of $\mid\theta\mid$ is in line with the evolution of $T_N$ [Fig. 4(c)] in Fe$_{1-x}$Co$_x$Sb$_2$S$_4$ ($x$ = 0, 0.1, and 0.2). Then the ratio $f = \mid\theta\mid/T_N$ could be calculated, which is about 8.63 for $x$ = 0 and decreases to 5.26 for $x$ = 0.1 and 5.07 for $x$ = 0.2, indicating moderate spin frustration in this system.\cite{Greedan, Dai} Moreover, there is an additional weak peak $T^*$ just above $T_N$ in the $d(\chi T)/dT$ vs $T$ curve, which is also confirmed by the heat capacity measurement (see the discussion below). The two-step magnetic transition was also observed in iron-based Fe$_{1+x}$Te with $x \geq 0.13$,\cite{Hu} benavidesite MnPb$_4$Sb$_6$S$_{14}$,\cite{Leone} and manganese-based MnBiS$_2$Cl.\cite{Brochard} Whereas FeSb$_2$S$_4$ M$\mathrm{\ddot{o}}$ssbauer experiment suggests a helicoidal-type AFM ground state with Fe$^{2+}$ moments parallel to the $ab$ plane and with a non-commensurate propagation vector along the $c$ axis, neutron diffraction studies are necessary to shed more details on the two-step transition.\cite{Bonville,BaoW} The hysteresis loops measured at $T$ = 2 K show a weak FM component at low fields, which might be caused by spin canting and/or magneto-crystalline anisotropy, as shown in Fig. 4(d). This FM component increases with increasing $x$ [inset of Fig. 4(d)], which is certified by the increase of coercive field $H_c$.

\begin{figure}
\centerline{\includegraphics[scale=0.9]{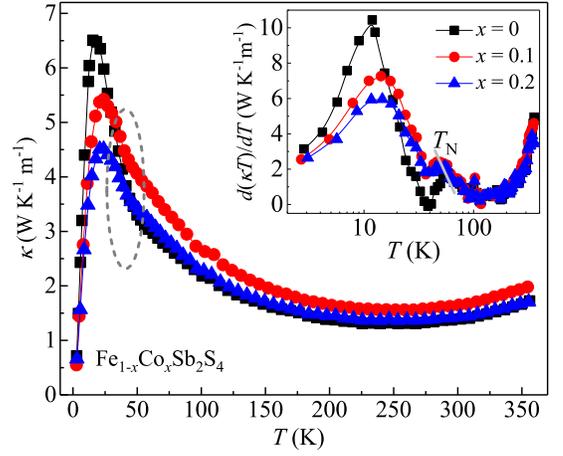}}
\caption{(Color online) Temperature-dependent thermal conductivity $\kappa(T)$ of Fe$_{1-x}$Co$_x$Sb$_2$S$_4$ ($x$ = 0, 0.1, and 0.2). Inset: $d(\kappa T)/dT$ vs $T$ curves. The solid line is guide to the eye.}
\label{kT}
\end{figure}

Figure 5 represents the temperature-dependent thermal conductivity $\kappa(T)$ of Fe$_{1-x}$Co$_x$Sb$_2$S$_4$ ($x$ = 0, 0.1, and 0.2). In general, $\kappa_{total} = \kappa_e + \kappa_{ph}$, consists of the electronic charge carrier part $\kappa_e$ and the phonon term $\kappa_{ph}$. The $\kappa_e$ part can be estimated from the Wiedemann-Franz law $\kappa_e/T = L_0/\rho$, where $L_0$ = 2.45 $\times$ 10$^{-8}$ W $\Omega$ K$^{-2}$ and $\rho$ is the measured electrical resistivity. The estimated $\kappa_e$ is less than 0.01 \% of $\kappa_{total}$ because of the large electrical resistivity of Fe$_{1-x}$Co$_x$Sb$_2$S$_4$ ($x$ = 0, 0.1, and 0.2), indicating a predominantly phonon contribution. At room temperature, the $\kappa(T)$ shows relatively low values of 1.39-1.64 W K$^{-1}$ m$^{-1}$, which could be contributed to the combination of low crystal symmetry and complex structure and chemical composition with heavy element Sb. Moreover, the $\kappa(T)$ shows weak temperature dependence above 150 K. With decreasing temperature, the observed increase in $\kappa(T)$ is consistent with a gradual freezing of phonon umklapp processes, and a typical phonon peak was observed around 20 K. With Co doping, the phonon peak of $x$ = 0 is about 6.5 W K$^{-1}$ m$^{-1}$ and it is suppressed significantly to 5.4 and 4.5 W K$^{-1}$ m$^{-1}$ for $x$ = 0.1 and 0.2, respectively. The suppression of $\kappa(T)$ should reflect enhanced phonon scattering, which is in general realized by grain boundary, point defects, carrier-phonon scattering, and phonon Umklapp scattering.\cite{Nolas, Callaway, Glassbrenner} The carrier concentrations in our samples are very low and the boundary scattering and Umklapp process should not vary significantly by replacing small amount of Fe with Co. Therefore the suppression of $\kappa(T)$ should be mostly contributed by the Fe/Co doping disorder enhanced point defects scattering. Most importantly, a notable hump feature was observed in the $\kappa(T)$ curves in addition to the phonon peaks. As shown in the inset of Fig. 5, the $d(\kappa T)/dT$ vs $T$ curves exhibit weak kinks around 50 K, of which the temperature slightly decreases with Co doping level $x$. The hump of $\kappa(T)$ is in good agreement with the observed magnetic transitions, indicating strong spin-lattice coupling in Fe$_{1-x}$Co$_x$Sb$_2$S$_4$.

\begin{figure}
\centerline{\includegraphics[scale=0.9]{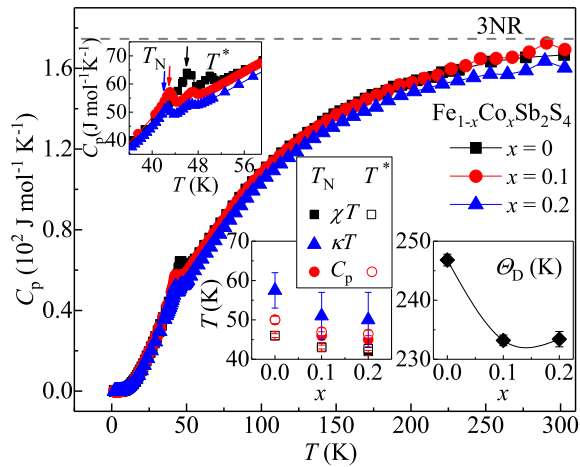}}
\caption{(Color online) Temperature-dependent heat capacity of Fe$_{1-x}$Co$_x$Sb$_2$S$_4$ ($x$ = 0, 0.1, and 0.2). Insets: The enlargement of the specific-heat anomaly between 36 K and 60 K, and the evolution of transition temperatures ($T_N$ and $T^*$) and Debye temperature ($\Theta_D$) as a function of Co content $x$.}
\label{CpT}
\end{figure}

The specific heat $C_p(T)$ of Fe$_{1-x}$Co$_x$Sb$_2$S$_4$ ($x$ = 0, 0.1, and 0.2) (Fig. 6) approaches the value of 3NR at room temperature, where N is the atomic number in the chemical formula (N = 7) and R is the gas constant (R = 8.314 J mol$^{-1}$ K$^{-1}$), consistent with the Dulong-Petit law. By neglecting the magnon contribution at low temperatures, the specific heat can be separated into the electronic and phonon parts, $C_p(T) = \gamma T + \beta T^3$. By fitting the $C_p(T)$ data below $T$ = 6 K, the obtained Sommerfeld electronic specific-heat coefficient $\gamma$ is less than 0.005 J mol$^{-1}$ K$^{-2}$, in line with its insulating ground state. For FeSb$_2$S$_4$ ($x$ = 0), the derived Debye temperature $\Theta_D$ = 247(1) K from $\beta$ = 0.90(1) mJ mol$^{-1}$ K$^{-4}$ using the equation $\Theta_D = [12\pi^4NR/(5\beta)]^\frac{1}{3}$ slightly decreases to 233(1) K for $x$ = 0.2. The enlargement of the specific-heat anomaly between 36 K and 60 K shows an obvious $\lambda$-type peak at $T_N$ = 46.0(5) K for $x$ = 0, corresponding to the formation of long-range AFM ordering, as well as an additional small peak at $T^*$ = 50(1)K, in good agreement with the magnetic transition observed in the susceptibility curve. This could indicate the subtle magnetostructural effects at the AFM transition, similar to Fe$_{1+y}$Te, calling for re-examination of the low temperature structure of FeSb$_{2}$S$_{4}$.\cite{Zaliznyak} The evolution of $T_N$ and $T^*$ with $x$ from different methods is finally summarized in the inset of Fig. 6.

\begin{figure}
\centerline{\includegraphics[scale=1]{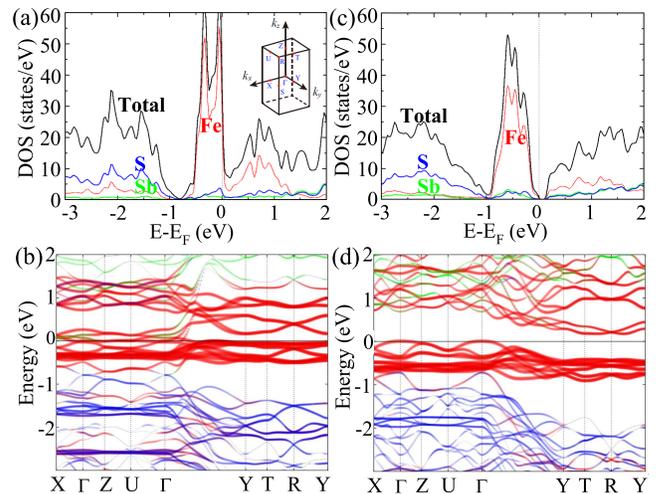}}
\caption{(Color online) Density of states and band structure of FeSb$_2$S$_4$ using experimental values (a,b) and DFT-relaxed lattice parameters (c,d). The states with Fe $d$, S $p$, and Sb $p$ character are denoted by thick red, medium blue, and thin green lines, respectively. Inset in (a) shows the sketch of the FeSb$_2$S$_4$ Brillouin zone.}
\label{DFT}
\end{figure}

To give a better description of the experimental data, we also calculated the band structure of a simple non-magnetic FeSb$_2$S$_4$ using experimental lattice parameters. First-principles calculations based on density functional theory demonstrate the dominance of Fe $3d$ states near the Fermi level and in partial density of states, as shown in Fig. 7. It is of interest to note that the experimental lattice parameters result in a metallic ground state within our theoretical framework. Since the standard GGA functional tends to underestimate band gaps of semiconductors, the modified Becke-John (mBJ) exchange potential was also utilized to verify the bulk band gap.\cite{TranF} However, the metallic ground state is robust even in the mBJ exchange potential. An indirect energy gap $\Delta$ = 158 meV opens up [Fig. 7(d)] only with fully relaxed structure where $a$ = 11.274 {\AA}, $b$ = 3.636 {\AA}, $c$ = 13.821 {\AA}, which are smaller than the experimental values [$a$ = 11.385(2) {\AA}, $b$ = 3.765(2) {\AA}, $c$ = 14.147(2) {\AA}]. This suggests that insight into the low-temperature crystal and magnetic structure is of interest.

\section{CONCLUSIONS}

Our study has demonstrated the polaronic nature of electronic transport in the magnetic semiconductor alloys Fe$_{1-x}$Co$_x$Sb$_2$S$_4$ ($x$ = 0, 0.1, and 0.2), based on the large discrepancy between activation energy for conductivity $E_\rho$ (146 $\sim$ 270 meV) and for thermopower $E_S$ (47 $\sim$ 108 meV). Bulk magnetization and heat capacity of FeSb$_2$S$_4$ exhibit two-step magnetic transitions with possible canted AFM ground state. The transition temperatures ($T_N$ and $T^*$) slightly decrease with increase Co doping level $x$. The magnetic transitions are also observed in the thermal conductivity measurement, demonstrating not only strong spin-lattice coupling but also thermal conductivity values much smaller from the values found in iron diantimonide. Even though the thermopower $S$ of FeSb$_{2}$S$_{4}$ is smaller when compared to FeSb$_{2}$, Fe$_{1-x}$Co$_x$Sb$_2$S$_4$ shows increase of thermopower with $x$. Given the similarity of its crystal structure to marcasites such as FeSb$_{2}$ but also a ternary chemical formula that offers additional tunability when compared to it, further anion-substitutions might enhance its thermoelectric performance.

\section*{Acknowledgements}
This paper was supported by the U.S. Department of Energy (DOE), Office of Science, Basic Energy Sciences as a part of the Computational Materials Science Program. This research used the 8-ID (ISS) beamline of the National Synchrotron Light Source II, a U.S. DOE Office of Science User Facility operated for the DOE Office of Science by Brookhaven National Laboratory under Contract No. DE-SC0012704.

\end{document}